\documentclass[aps,amsmath,twocolumn,prb,showpacs]{revtex4}

\usepackage{graphicx}
\usepackage{color}

\begin{document}

%\preprint{}

\title{Partially suppressed long-range order in the Bose-Einstein condensation of
polaritons}

\author{D. Sarchi}
\email[]{davide.sarchi@epfl.ch}
\affiliation{Institute of Theoretical Physics, Ecole Polytechnique F\'ed\'erale de Lausanne EPFL, CH-1015 Lausanne, Switzerland}
\author{V. Savona}
\affiliation{Institute of Theoretical Physics, Ecole Polytechnique F\'ed\'erale de Lausanne EPFL, CH-1015 Lausanne, Switzerland}

\date{\today}

\begin{abstract}
We adopt a kinetic theory of polariton non-equilibrium
Bose-Einstein condensation, to describe the formation of
off-diagonal long-range order. The theory accounts properly for
the dominant role of quantum fluctuations in the condensate. In
realistic situations with optical excitation at high energy, it
predicts a significant depletion of the condensate caused by
long-wavelength fluctuations. As a consequence, the one-body
density matrix in space displays a partially suppressed
long-range order and a pronounced dependence on the finite size
of the system.
\end{abstract}


\maketitle                   % Produces the title.

\section{Introduction}
Bose-Einstein condensation (BEC) is one of the most remarkable
manifestations of quantum mechanics at the macroscopic
scale.\cite{pines66} {The key feature of BEC of an interacting
Bose gas is the formation of off-diagonal long range order
(ODLRO),\cite{bloch00,penrose56} i.e. the fact that the spatial
correlation $g^{(1)}({\bf r},{\bf r}^\prime)=n({\bf r},{\bf
r}^\prime)/\sqrt{n({\bf r})n({\bf r}^\prime)}$ (where $n({\bf
r},{\bf r}^\prime)$ is the one-body density matrix, while $n({\bf
r})$ is the particle density) extends over the whole system
size}. The BEC mechanism is the key to understand
superconductivity and superfluidity,\cite{pines66} and was the
object of renewed interest following its recent discovery in
diluted alkali atoms.\cite{pita03} Another candidate system for
the observation of BEC is that of excitons\cite{griffin95} or
exciton-polaritons\cite{snoke02} in insulating crystals.
Recently, in particular, many
theoretical~\cite{marchetti06,laussy04,doan05,schwendimann06} and
experimental~\cite{deng02,deng03,richard05,kasprzak06} efforts
have been devoted to BEC of microcavity polaritons. Due to the strong mutual polariton interactions, the 
system is expected to deviate significantly from the ideal Bose
gas picture. Interactions are responsible of quantum fluctuations, namely of the emergence of collective eigenmodes that differ from the single-particle states.\cite{pita03} 
Quantum fluctuations in the polariton gas are expected to be
more relevant than in a diluted atomic gas. This will lead to a
depletion of the condensate that can be very important, reminding
of the prototypical case of $^4$He in which a condensate fraction
of less than 10\% at equilibrium is achieved. {Deviations from thermal 
equilibrium might still enhance this tendency. As an example, theoretical predictions for 
bulk polaritons suggest that an initial condensate population can be
totally depleted by quantum fluctuations within a sufficiently long time~\cite{beloussov96}. 
In the present case, a full depletion is only prevented by the short 
polariton lifetime, allowing to a considerable fraction of the particles in the condensate 
to undergo radiative recombination before scattering to the excited states}. Another 
indication of the importance of quantum fluctuations comes from the 
measurement\cite{deng02} of the
second-order time-correlation function at zero delay $g^{(2)}(0)$. There, quantum 
fluctuations might explain the large deviation
from the quantum limit ($g^{(2)}(0)=1$), observed far above the condensation
threshold. The role played by quantum fluctuations
is even more important in the light of the two-dimensional nature
of microcavity polaritons. For a two-dimensional system in the
thermodynamic limit, as stated by the Hohenberg-Mermin-Wagner
theorem,~\cite{hohenberg67} the long-wavelength fluctuations
diverge, and ODLRO cannot arise. However condensation is
prohibited only in infinitely extended systems. Realistic
polariton systems, based both on III-V~\cite{LangbeinPRL02} and
on II-VI~\cite{richard05} semiconductors, exhibit disorder that
tends to localize the lowest polariton levels over a few tens of
$\mu$m~\cite{LangbeinPRL02}. In these situations, it was
rigorously proved that the discrete energy spectrum, arising
either from quantum confinement in a finite
system~\cite{lauwers03} or from disorder induced polariton
localization,~\cite{lenoble04} allows condensation in a thermal
equilibrium situation, although the occurrence of ODLRO could be
inhibited, depending on the localization length and on the
disorder amplitude.~\cite{lenoble04} Very recently, a direct
measurement of the spatial correlation function in a II-VI
semiconductor based microcavity~\cite{kasprzak06}, provided a
striking experimental signature of polariton condensation with
formation of ODLRO.

In order to experimentally assess polariton BEC, it is important
to predict how ODLRO manifests itself in a non-equilibrium
situation, in a localized geometry, and in presence of quantum
fluctuations. To this purpose, a field-theoretical approach is
required for a proper description of the relaxation kinetics and
of the quantum fluctuations in presence of mutual interaction.

In this work we develop a kinetic theory of polaritons subject to
mutual interaction, in which the field dynamics of collective
excitations is treated self-consistently along with the
condensation kinetics. We start from a number-conserving Bogolubov approach \cite{zoller98,castin98}
that describes the collective modes of a Bose gas properly accounting for the number of
particles in the condensate. This is required in order to develop kinetic equations for the
description of condensate formation. We then derive a hierarchy of density matrix equations,
including polariton-phonon scattering via deformation-potential interaction and exciton-exciton
scattering in the exciton-like part of the lower-polariton branch. This latter mechanism is based
on the model recently developed by Porras {\em et al.} \cite{porras02}. The hierarchy is
truncated to include coupled equations for the populations in the lower polariton branch and for
the two-particle correlations between the condensate and the excitations. For the truncation,
we assume that higher-order correlations evolve much faster than the relaxation dynamics. The
kinetic equations obtained in this way are solved numerically, assuming a steady-state pump at
high energy within the exciton-like part of the polariton branch. The solution is carried out by
accounting self-consistently for the density-dependent Bogolubov spectrum of the collective
excitations of the polariton gas. We show how this model predicts a
dominant effect of quantum fluctuations that result in a significant
condensate depletion under typical excitation conditions. In
particular, we discuss the role of quantum confinement in a
system of finite size and show how ODLRO manifest itself in
typical experimental conditions. Our results provide a clear explanation of the
partial suppression of ODLRO that characterizes the experimental findings~\cite{kasprzak06}.

The paper is organized as follows. In Section II we describe in detail the theoretical framework of the present analysis. Section III
is devoted to the presentation of the numerical solution of the kinetic equations and to the
discussion of the results in the light of recent polariton BEC experiments. In Section IV we
present our conclusions.

\section{Theory}
We consider the polariton in the lower branch of the dispersion
as a quasi-particle in two dimensions,\cite{note} described by
the Bose field $\hat{p}_{k}$, obeying
$[\hat{p}_{k},\hat{p}^{\dagger}_{k'}]=\delta_{kk'}$. The lower
polariton Hamiltonian in presence of Coulomb and polariton-phonon
scattering is \cite{ciuti03}
\begin{eqnarray}
H&=&H_{0}+\frac{1}{2}\sum_{kk'q}v^{(q)}_{kk'}\hat{p}^{\dag}_{k+q}\hat{p}^{\dag}_{k'-q}\hat{p}_{k'}\hat{p}_{k}\nonumber\\
&+&\sum_{kk'q}g^{(q)}_{kk'}(b^{\dagger}_{q}+b_{-q})(\hat{p}^{\dag}_{k}\hat{p}_{k'}+\hat{p}^{\dag}_{k'}\hat{p}_{k}),
\label{eq:Htot}
\end{eqnarray}
where
$H_0=\sum_{k}\hbar\omega_{k}\hat{p}^{\dag}_{k}\hat{p}_{k}+\sum_{q}\hbar\omega_{q}b^{\dagger}_{q}b_{q}$
is the free Hamiltonian for polaritons and phonons. The quantity
$v^{(q)}_{kk'}$ arises from the Coulomb interaction between
excitons $v_{XX}$ and from the oscillator strength saturation due
to Pauli exclusion $v_{sat}$~\cite{Rochat00,ben01,Okumura2002},
{as
\begin{eqnarray}
v^{(q)}_{kk'}&=&v_{XX}X_{k+q}X_{k'-q}X_{k}X_{k'}+\nonumber \\
&& +v_{sat}X_{k'-q}(C_{k+q}X_{k}+X_{k+q}C_{k})X_{k'}.
\label{eq:ppint}
\end{eqnarray}
Here the $X_{k},C_{k}$ are the Hopfield coefficients representing
the excitonic and the photonic weights respectively, in the lower
polariton field, i.e.
$\hat{p}_{k}=X_{k}\hat{B}_{k}+C_{k}\hat{A}_k$. Here $\hat{B}_k$
and $\hat{A}_k$ are the exciton and the photon destruction
operators respectively~\cite{ciuti03}. In Eq.(\ref{eq:ppint}) we
consider the interaction matrix elements in the small-momentum
limit (consistently with the restriction of the analysis to the
lower polariton branch of the
dispersion)~\cite{Rochat00,porras02,doan05}, i.e.
\begin{eqnarray}
v_{XX}=\frac{6E_{b}a_{0}^2}{A} \nonumber \\
v_{sat}=-\frac{\hbar\Omega_R}{n_{sat}A}, \label{eq:vme}
\end{eqnarray}
where $E_{b}$ is the exciton binding energy, $a_0$ is the exciton
Bohr radius, $2 \hbar\Omega_R$ is the microcavity vacuum-field
Rabi splitting, $n_{sat}=7/16\pi a_0^2$ is the exciton saturation
density~\cite{Rochat00,Schmitt-Rink1985} and $A$ is the system
quantization area. The quantity $g^{(q)}_{kk'}$ describes the
deformation potential interaction of polaritons with acoustic
phonons \cite{tassone97,doan05} and reads
\begin{eqnarray}
g^{(q)}_{kk'}&=&i\sqrt{\frac{\hbar |q|}{2 \rho A L_z
u}}X_{k}X_{k'}\times \nonumber \\
&& \times
[a_eI^{||}_e(|q|)I^{\bot}_e(q_z)-a_hI^{||}_h(|q|)I^{\bot}_h(q_z)].
\label{eq:gphon}
\end{eqnarray}
Here, $\rho$ and $u$ are the density and the longitudinal sound
velocity in the semiconductor, respectively, $L_z$ is the quantum
well width and $a_{e(k)}$ are the electron(hole) deformation
potentials. The terms $I^{||}_{e(h)}(|q|)$ and
$I^{\bot}_{e(h)}(q_z)$ are the superposition integrals of the
exciton envelope function with the phonon wave function (plane
wave) in the in-plane and $z$-directions, respectively, and read
\begin{eqnarray}
I^{||}_{e(h)}(|q|)=[1+(a_0q_zm_e(h)/2(m_e+m_h))^2]^{-3/2}\nonumber \\
I^{\bot}_{e(h)}(q_z)=\int dz |f_{e(h)}(z)|^2e^{iq_z z},
\label{eq:intsup}
\end{eqnarray}
where $m_{e(h)}$ is the electron (hole) mass while $f_{e(h)}(z)$
are the electron(hole) envelope functions in the growth
direction, according to the exciton envelope function
picture~\cite{piermarocchi96}}.

For a kinetic description, a number-conserving approach
\cite{zoller98,castin98} is required. This formalism allows
treating in a self-consistent way the field dynamics and the
population kinetics, and describes correctly the condensate in-
and out-scattering rates. In the number-conserving approach, the
polariton field is expressed as
\begin{equation}
\hat{p}_k= P_{k}\hat{a}+\tilde{p}_{k},
\end{equation}
i.e. the sum of a condensate operator $P_{k}\hat{a}$ and a
single-particle excitation operator
$\tilde{p}_{k}$.\cite{castin98} The condensate operator obeys
Bose commutation rules $[\hat{a},\hat{a}^{\dagger}]=1$. The
quantity $N_{c}=\langle \hat{a}^{\dagger} \hat{a} \rangle$ is the
population of condensed particles, while $P_{k}$ represents the
condensate wave function in momentum space. The single-particle
excitation field $\tilde{p}_{k}$ is orthogonal to the wave
function of the condensate, i.e.
$\sum_{k}P^{*}_{k}\tilde{p}_{k}=0$, and obeys the modified Bose
commutation rule
$[\tilde{p}_k,\tilde{p}_{k'}^{\dagger}]=\delta_{kk'}-P_{k}P_{k'}^{*}$.
Using these definitions, the total population at momentum $k$ is
$N_{k}=\langle \hat{p}^{\dagger}_{k}\hat{p}_{k}\rangle =|P
_{k}|^{2} N_{c} + \tilde{N}_{k}$, where $\tilde{N}_{k}=\langle
\tilde{p}^{\dagger}_{k}\tilde{p}_{k}\rangle$ is the population of
non-condensed particles. Within the number conserving approach, a
quantum fluctuation is defined by the operator\cite{castin98}
\begin{equation}
\hat{\Lambda}^{\dagger}_k\equiv\frac{1}{\sqrt{N}}\hat{a}\tilde{p}^{\dagger}_k\,
\label{eq:bogo}
\end{equation}
that promotes a particle from the condensate to the excited
states. $N$ is the total number of particles. This field can be
formally written via a Bogolubov transformation as
$\hat{\Lambda}_k=U_k\hat{\alpha}_k+V^*_{-k}\hat{\alpha}^{\dagger}_{-k}$,
where $U_k$ and $V^*_{-k}$ are modal functions, and
$\hat{\alpha}_k$ are Bose operator describing the collective
excitations at energy
$E_k$.\cite{pita03,castin98}.

\begin{figure}[ht]
\includegraphics[width=.47 \textwidth]{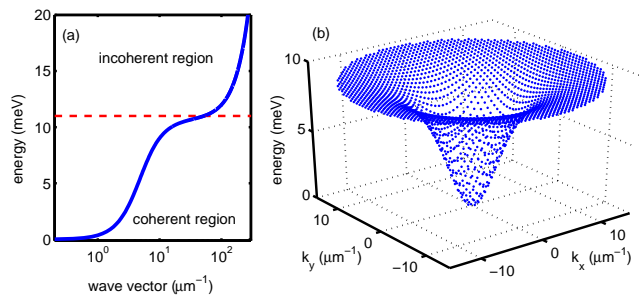}
\caption{(a) Energy-momentum lower-polariton dispersion. Notice
the logarithmic horizontal scale. (b) Energy-momentum plot of the
discrete lower-polariton states in the coherent region, as used
in the simulations for $A=100~\mu\mbox{m}^2$.} \label{fig1}
\end{figure}

Our model is based on two key-assumptions. First, we separate the
single-particle energy spectrum into a lower energy coherent part
and an incoherent part at higher energies. This is depicted in
Fig. \ref{fig1} (a) for the typical energy-momentum dispersion of
the lower polariton branch. {In a condensate, collective Bogolubov excitations affect mostly the lower-energy part of the spectrum. As the energy becomes 
larger than the total interaction energy $vN_c$ (where $v$ is a measure of the interaction matrix element), $U_k\rightarrow1$ and $V_k\rightarrow0$, and the collective 
modes thus approach the single-particle states.
Hence, it is mostly in
the low-energy region that quantum fluctuations will affect the
condensate kinetics.} A
customary\cite{gardiner98} approximation consists in restricting the quantum kinetic
treatment to the coherent region, while the dynamics within
the incoherent region is modeled in terms of a simple Boltzmann population
kinetics.

The separation, for the lower polariton branch, naturally
coincides with that between the strong-coupling region and the
flat exciton-like part of the dispersion, as illustrated in Fig.
\ref{fig1}(a). The second approximation is made possible by the
remark that, given the large Coulomb scattering amplitude, the
field dynamics of collective Bogolubov excitations takes place
much faster than energy relaxation mechanisms, made slow by the
steep polariton energy-momentum dispersion that reduces the space
of final states available for scattering processes. We thus
assume that, on the timescale of the relaxation, a
quasi-stationary spectrum of collective Bogolubov excitations
arises, that evolves adiabatically and is computed
self-consistently at each time step in the kinetics, by means of
the Popov version of the Hartree-Fock-Bogolubov approximation
(HFB Popov). {This approximation is justified by the two
following arguments. First, in equilibrium conditions, it is
known that the HFB Popov approximation predicts collective
excitations in excellent agreement with the measured excitation
spectra of a weakly interacting Bose gas, when the temperature is
sufficiently low, i.e. $T\leq0.5 T_c$, where $T_c$ is the
critical temperature. Alternatively, this condition corresponds
to a density considerably above the critical
density.\cite{dodd98,liu04} Second, our present purpose is not to
determine the exact spectrum of the collective multipole
oscillations of the polariton gas close to the condensation
threshold. It is rather to estimate how the density-dependent
changes in the spectrum affect the relaxation dynamics and the
coherent scattering processes significantly above the
condensation threshold, when ODLRO becomes detectable. In
addition, close to threshold, the polariton relaxation dynamics
and the coherent scattering processes are expected to be only
marginally affected by the the details of the spectrum, because
the populations in the condensate and in the low-lying excited
states are small, as argued in Ref. \onlinecite{porras02}.}

The kinetics is described in terms of a density-matrix hierarchy
whose time-evolution is obtained from the Heisenberg equations of
motion. Coulomb interaction terms within the incoherent region
are treated consistently with the Boltzmann picture, as was done
by Porras {\em et al.} \cite{porras02}. In particular, we assume
that, in the incoherent region, the population of the
exciton-like polaritons is thermally distributed with an
effective temperature defined by $k_BT_x\equiv e_x/n_x$, where
$k_{B}$ is the Boltzmann constant, $n_x$ is the total particle
density and $e_x$ is the total energy density in the incoherent
region (as shown below, these two quantities are determined
self-consistently during relaxation, via Boltzmann equations).
Within this picture, the processes of two particles in the
incoherent region scattering into one particle in the incoherent
region and one in the coherent region, result in an effective
energy relaxation mechanism towards the bottom of the polariton
branch.
In the coherent region, on the other hand, we extend the
hierarchy of equations to the next order, thus including two-body
(four operator) correlations. The most important of these
two-body correlations entering our equations describes coherent
processes where two polaritons are scattered between the
condensate and the excited states within the coherent region. It
is given by
$\tilde{m}_k=N\langle\hat{\Lambda}_k\hat{\Lambda}_{-k}\rangle$ and
describes the main effect of quantum fluctuations. These
processes do not conserve energy and could not be described in
terms of Boltzmann equations for the populations of the
single-particle states. They are made possible only because, in a
condensed system, the actual eigenstates $\{|\nu\rangle\}$ are
collective modes and differ from the single-particle states
$\{|k\rangle\}$. As a consequence, the quantities
$\langle\nu|\hat{\Lambda}_k\hat{\Lambda}_{-k}|\nu\rangle$ are
finite. {As already pointed out, their amplitude is expected to decrease for
increasing energy $E_k$. We have checked that our model reproduces this behaviour, as is reported below in connection with
Fig.~\ref{fig:4}. This is proves that our approximation, consisting in the separation into
two energy regions, is consistent with the real behaviour of the
system. A full treatment in terms of quantum kinetics is made here prohibitive by the inclusion of a very large number of states in the exciton-like part of the dispersion, that 
is needed in order to model the initial high-energy excitation used in photoluminescence experiments. 
Models assuming an initial population in the lower-energy part of the spectrum\cite{beloussov96}, can instead treat the full spectrum consistently. In fact, in that case, 
the only relevant coherent scattering processes involve states with energy smaller
than the interaction energy $v N$ (where $v$ is the interaction
matrix element and $N$ is the initial condensate population, see e.g. Fig.~3 of Ref.~\onlinecite{beloussov96}).}

The polariton-phonon scattering is treated within a shifted-pole
Markov approximation, resulting in standard Boltzmann
contributions \cite{tassone97}. For the calculations, we assume a
finite-size homogeneous system of square shape and area $A$ with
periodic boundary conditions, resulting in spatially a uniform
condensate wave function, i.e.
$P_k=e^{i\phi}\delta_{k,0}$.\cite{leggett01} In a realistic
condensate, this assumption is valid everywhere, except within a
distance from the boundary equal to the healing length
$\xi=\hbar/\sqrt{M v n}$~\cite{pita03}. For polaritons, we find
$\xi\approx 1~\mu\mbox{m}$ for the estimated density threshold.
The confinement can model both finite size polariton
traps~\cite{EldaifAPL2006} and the situation close to a local
minimum of the disorder potential in extended
systems.\cite{richard05,LangbeinPRL02} The finite size results in
quantum confinement and thus in a discrete energy spectrum, with
a gap between ground and first excited state
$\Delta=\hbar^2(2\pi)^2/(M_{pol}A)$.\cite{doan05}

Within these prescriptions (see Appendix) the kinetic equations are:
\begin{eqnarray}
\dot{N}_c  &=&  - \gamma_0 N_{c} + \dot{N}_{c}\left.\right|_{ph} +
\dot{N}_{c}\left.\right|_{XX} + \frac{2}{\hbar}\sum_{k}v^{(k)}_{k,-k}{\mbox{Im}} \{\tilde{m}_k\} \nonumber \\
\dot{\tilde {N}}_k  &=&  - \gamma_k \tilde{N}_k +
\dot{\tilde{N}}_{k}\left.\right|_{ph} +
\dot{\tilde{N}}_{k}\left.\right|_{XX} -
\frac{2}{\hbar}v^{(k)}_{k,-k}{\mbox{Im}} \{\tilde{m}_k\} \nonumber \\
\dot{\tilde m}_k
&=&-2\left[\gamma_0+i\omega_{k}+\frac{i}{\hbar}v^{(0)}_{k,0}(N_c-\tilde{N}_k-5/2)\right]\tilde{m}_{k}\nonumber\\
&-&
\frac{i}{\hbar}\left[\sum_{q}v^{(k-q)}_{q,-q} \tilde{m}_q - 2 v^{(k)}_{k,-k} N_c(N_c-1)\right](1+2\tilde{N}_k)\nonumber\\
&+&
2\frac{i}{\hbar}(1+2N_c)\sum_{q}v^{(q)}_{q,-q}\langle\tilde{p}^{\dagger}_q
\tilde{p}^{\dagger}_{-q}\tilde{p}_k\tilde{p}_{-k}\rangle \nonumber \\
\dot{n}_{x}&=& -\gamma_x n_x + \dot{n}_{x}\left.\right|_{ph} +
\dot{n}_{x}\left.\right|_{XX} + f. \label{eq:tot}
\end{eqnarray}
The $\gamma_k=\gamma_c |C_k|^2$ is the polariton radiative
lifetime, $|C_k|^2$ being the photon fraction in the polariton
state, $\gamma_c$ the cavity photon lifetime, and $\gamma_x$ the
exciton lifetime. The quantity $f$ denotes the pump intensity
producing a population in the incoherent region. The suffixes
``$ph$'' and ``$XX$'' denote the Boltzmann scattering terms for
polariton-phonon scattering~\cite{tassone97} and for
exciton-exciton scattering in the incoherent region
\cite{porras02}.

In the equation for $\tilde{m}_k$ some simplifications were
introduced. First, we have fully neglected the contributions due
to both energy relaxation mechanisms. This is consistent with our
adiabatic assumption, as the dynamics of the correlation
$\tilde{m}_k$ reflects the time-evolution of collective
excitations, taking place on a much faster timescale than
relaxation. Second, the higher order three-body correlations,
arising as the next level of the correlation hierarchy, have been
factored into products of one- and two-body correlations.
Furthermore, we assume the identity $\langle
\hat{a}^{\dagger}\hat{a}^{\dagger}\hat{a}\hat{a}
\rangle=N_c(N_c-1)$ to hold, as expected for a macroscopic
condensate occupation, $N_c\gg1$. Third, always according to the
adiabatic approximation, the two-body correlation function {\em
between} condensate excitations, $\langle\tilde{p}^{\dagger}_q
\tilde{p}^{\dagger}_{-q}\tilde{p}_k\tilde{p}_{-k}\rangle$ is
evaluated in a quasi-stationary limit. It can thus be expressed in
terms of the modal functions $U_k$ and $V^*_{-k}$ obtained by
diagonalizing the stationary Bogolubov problem at each time step
in the kinetics. Starting from the dynamical equation
$\dot{\hat{\Lambda}}_k=(\omega_k+v^{(0)}_{k,0}\xi_k)\hat{\Lambda}_k+v^{(k)}_{k
,k}\xi_k\hat{\Lambda}^{\dagger}_k$, with
$\xi_k=(N_c-\tilde{N}_k)/N$, we derive the actual eigenvalues
$E_k=[(\omega_k+v^{(0)}_{k,0}\xi_k)^2-
(v^{(k)}_{k,k}\xi_k)^2]^{1/2}$, while the modal functions are
given by
\begin{equation}
\left|V_k\right|^2=\xi_k\frac{\left[E_k-(\omega_k+v^{(0)}_{k,0}\xi_k)\right]^2}{(v^{(k)}_{k,k}\xi_k)^2_k-\left[E_k-(\omega_k+v^{(0)}_{k,0}\xi_k)\right]^2}
\label{eq:bogfac}
\end{equation}
and the normalization $|U_k|^2-|V_k|^2=\xi_k$. In this limit (see Appendix) we
can replace in (\ref{eq:tot}) the following expression
\begin{eqnarray}
&&\langle
\tilde{p}^{\dagger}_{q}\tilde{p}^{\dagger}_{-q}\tilde{p}_{k}\tilde{p}_{-k}\rangle
\simeq  \Upsilon(N) \left[\left|\sum_k U_k V^*_k(1+2\bar{N}_k)\right|^2\right. \nonumber \\
&&+ \left. \sum_k 2\chi_k \bar{N}_k \left(\chi_k \bar{N}_k+2\left|
V_k \right|^2\right)+2\left| V_k \right|^4 \right],
\label{eq:2corr_bog}
\end{eqnarray}
where $\Upsilon(N)=N^2[(N_c+1)(N_c+2)]^{-1}$ and
$\chi_k=\xi_k+2\left| V_k \right|^2$.
{$\bar{N}_k=\langle \hat{\alpha}^{\dagger}_k
\hat{\alpha}_k \rangle$ is the population of the Bogolubov modes,
which can be expressed in terms of the single-particle
population via the exact relation $\tilde{N}_k = (N/(N_c+1))\times[(|U_k|^2 +
|V_k|^2)\bar{N}_k + |V_k|^2]$.} This finally brings to a closed
set of kinetic equations for the amplitudes $\tilde{m_k}$, the
populations $N_c$, $\tilde{N}_k$, and the total density $n_x$ in
the incoherent region.

{Before discussing the numerical solution of Eqs. (\ref{eq:tot}),
we present the detailed expressions of the Boltzmann terms
entering these equations. Their expressions are formally
identical to the corresponding terms derived in
Ref. \onlinecite{doan05} (for polariton-phonon scattering) and in
Ref. \onlinecite{porras02} (for exciton-exciton scattering), with an
important difference: in the present case, the actual HFB Popov
spectrum $E_k$ replaces the non-interacting single-particle
spectrum. Here we report the expressions of these terms, the
details of the derivation being described in Refs.
\cite{porras02,doan05}.
\begin{eqnarray}
\dot{N}_{k}\left.\right|_{ph}&=&A\left[W^{ph}_{x\rightarrow k}
n_{x} (1+N_k)-W^{ph}_{k\rightarrow x} N_{k}
(\eta+\frac{N_{k'}}{A})\right] + \nonumber \\
&+&\sum_{k'\in U_{coh}}\left[W^{ph}_{k'\rightarrow k} N_{k'}
(1+N_k)\right. + \nonumber \\
&&-\left.W^{ph}_{k\rightarrow k'} N_{k} (1+N_{k'})\right],
\label{eq:phonterms}
\end{eqnarray}
\begin{equation}
\dot{N}_{k}\left.\right|_{XX}=W^{XX}_{x\rightarrow k} n^2_{x}
(1+N_k)-W^{XX}_{k\rightarrow x} n_x N_{k},  \label{eq:xxterms}
\end{equation}
\begin{equation}
\dot{n}_{x}\left.\right|_{ph}=\sum_{k\in
U_{coh}}\left[W^{ph}_{k\rightarrow x} N_{k}
(\eta+n_x)-W^{ph}_{x\rightarrow k} n_x (1+N_{k})\right]
\label{eq:xphterm}
\end{equation}
and
\begin{equation}
\dot{n}_{x}\left.\right|_{XX}=-\frac{1}{A}\sum_{k\in
U_{coh}}\left[W^{XX}_{x\rightarrow k} n_x^2 (1+N_{k})
-W^{XX}_{k\rightarrow x} n_x N_{k}\right], \label{eq:xXXterm}
\end{equation}
where $\eta A \equiv [(m_e+m_h)k_BT_x]A/2\pi\hbar^2$ is the
number of states in the incoherent region, and $U_{coh}$ denotes
the coherent region. Here, $N_{k}=\tilde{N}_k$ for $k\neq 0$
and $N_{0}=N_c$. Note that in
Eqs.~(\ref{eq:phonterms}-\ref{eq:xXXterm}) we use the fact that
$n_x/\eta\ll 1$ and $k\ll k_x$, where $k_x$ is the averaged
momentum of the exciton distribution in the incoherent region. The rates appearing in
Eq.~(\ref{eq:phonterms}-\ref{eq:xXXterm}) are given by
\begin{eqnarray}
W_{k \to k'}^{ph} &&=
\frac{2L_z}{\hbar}\left|g^{\bar{q}}_{k,k'}\right|^2
\frac{\left|E_{k}-E_{k'}\right|}{(\hbar u)^2
q_z}\left[n^{ph}_{kk'}+ \theta \left(E_k -
E_{k'}\right) \right], \nonumber \\
&& |\bar{q}|=\sqrt{|k-k'|^{2}+\bar{q}^2_{z}}=|E_k-E_{k'}|/\hbar u,
\label{eq:wph}
\end{eqnarray}
where $n^{ph}_{kk'}$ is the population of phonons having energy
$E=E_k-E_{k'}$;
\begin{equation}
W_{x \to k}^{XX}  = \frac{2\pi}{\hbar k_B T_x }\left|v_{k_x ,k_x
}^{\left(k_x \right)}\right|^2 A^2 e^{\left( E_k  - E_{k_x}
\right)/k_BT_x}; \label{eq:Wxk}
\end{equation}
\begin{equation}
W_{k \to x}^{XX}  = \frac{(m_e+m_h)}{\hbar ^3}\left|v_{k_x ,k_x
}^{\left(k_x\right)} \right|^2 A^2 e^{2\left(E_k-E_{k_x}
\right)/k_B T_x}. \label{eq:Wkx}
\end{equation}
In all these expressions, consistently with our assumption the spectrum in the incoherent region
$E_{k_x}$ is the single-particle spectrum, only accounting for the overall density-induced blue shift of the polariton branch. Besides the set of equations (\ref{eq:tot}), 
the model is
completed by the introduction of an equation describing the
evolution of the energy density in the incoherent region
$e_x$. Following the procedure of Ref. \onlinecite{porras02}, this
equation reads
\begin{eqnarray}
\dot{e}_x&=&-\sum_{k\in U_{coh}}\frac{E_k}{A}\left[W_{x \rightarrow k}^{XX}n_x^2 \left(1 + \tilde N_k\right)-W_{k \rightarrow x}^{XX} n_x N_k\right]\nonumber \\
&-& \gamma_x \left(k_B T_x\right) n_x + \left(k_B
T_L\right)f-w^{ph}. \label{eq:ex}
\end{eqnarray}
Here, the first term represents the heating of the
incoherent population, produced by the exciton-exciton scattering
process and imposed by energy conservation; the second
term is the cooling due to the exciton radiative recombination;
the third term is originated by the assumption that the
incoherent population is created at the lattice temperature $T_L$;
the fourth term represents the cooling induced by exciton-phonon
coupling and it is evaluated as in Eq.~(21) of
Ref. \onlinecite{porras02}.}

\section{Numerical results}
\begin{figure}[ht]
\includegraphics[width=.43 \textwidth]{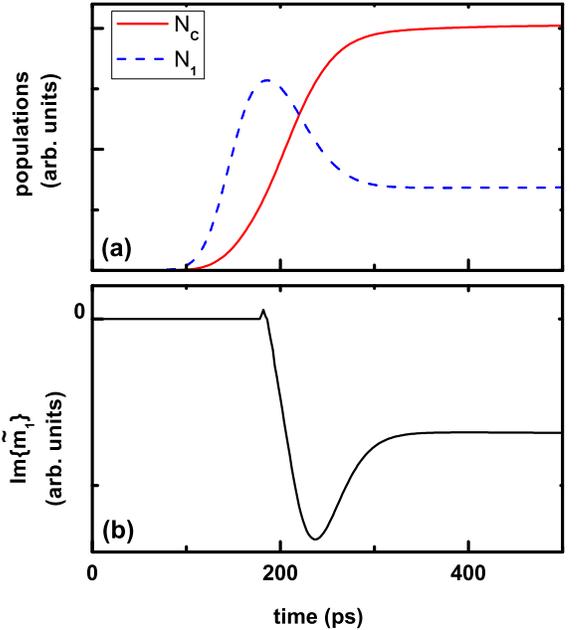}
\caption{Time dependent results for the populations of the
condensate state $N_c$ and of the first excited state $N_{1}$ and
for the imaginary part of the coherent scattering amplitude
$\tilde{m}_{1}$, for a given value of the excitation pump $f$
above condensation threshold. After an initial transient, all the
quantities reach a stationary value.} \label{fig:2}
\end{figure}

\begin{figure}[ht]
\includegraphics[width=.43 \textwidth]{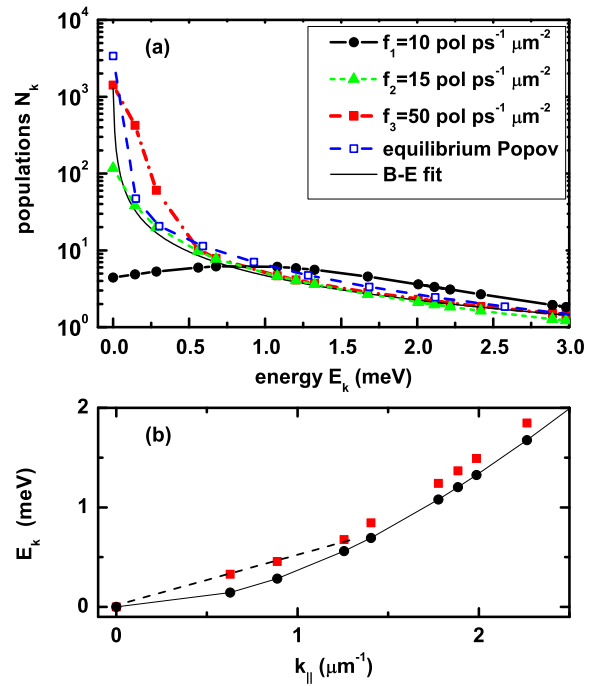} 
\caption{(a)
Steady state populations for increasing pump intensity $f$. Open
squares: equilibrium HFB Popov solution for (n,T) corresponding to
the steady-state solution for $f=f_3$. Thin line: B-E
distribution fitted to the high-energy tail and to the condensate
population for $f=f_3$. (b) Energy dispersion below and above
threshold (same legend as above). The dashed line is a guide to
the eye to highlight the linear part of the
dispersion.}\label{fig:3}
\end{figure}

Numerical solutions assuming a steady state pump have been
computed in the time domain. We assume parameter values relative
to the CdTe microcavity of Ref. \onlinecite{kasprzak06}, with Rabi
splitting $2\hbar\Omega_{R}=26~\mbox{meV}$ and cavity
photon-exciton detuning $\delta=5~\mbox{meV}$, at the lattice
temperature $T=10~\mbox{K}$. The quantization area is assumed
everywhere $A=100~{\mu}\mbox{m}^{2}$, unless specified,
consistent with estimates of polariton localization
length,~\cite{richard05} and gives rise to the discrete set of
polariton states plotted in Fig. \ref{fig1}(b).
{For this system we obtain from Eq. (\ref{eq:vme})
$v_{XX}=3.3\times 10^{-5}~\mbox{meV}$, $v_{sat}=-0.5\times
10^{-5}~\mbox{meV}$ and the resulting polariton-polariton
interaction matrix element at zero momentum is $v^{(0)}_{0,0}=6\times
10^{-5}~\mbox{meV}$}. The solutions always display steady-state
long-time values after an initial transient, as shown in
Fig.~(\ref{fig:2}). Here we also notice that, during the early
stages of the condensate growth, the scattering processes favor
condensation, through the positive values taken by
${\mbox{Im}}\{\tilde{m}_{k}\}$. immediately
afterwards, when a large condensate population is reached, the
quantity ${\mbox{Im}}\{\tilde{m}_{k}\}$
changes sign and coherent scattering terms start depleting the
condensate.
In Fig.~\ref{fig:3}~(a), we plot the steady-state populations per
state, for varying pump intensity. A pump threshold is found at
about $f=12~\mbox{ps}^{-1}~\mu\mbox{m}^{-2}$,
{corresponding to a polariton density $n=N/A\simeq
10~\mu m^{-2}$ and a total exciton density $n_x\simeq
100~\mu\mbox{m}^{-2}$ in the incoherent region. Let us remind that the system studied in
Ref. \onlinecite{kasprzak06} is composed by $16$ quantum wells. As the
polariton modes are expected to have similar amplitudes at
all the quantum well positions, in the experimental literature the exciton
density per quantum well is usually estimated by simply dividing
the total exciton density by the number of quantum
wells~\cite{deng03,richard05}. In the present case, following
this prescription, the exciton density per quantum well
is 7 $\mu\mbox{m}^{-2}$, i.e. two orders of magnitude lower than the exciton saturation density
in CdTe $n_{sat}\simeq 5\times10^{3}~\mu\mbox{m}^{-2}$. This confirms the validity of our
description of polaritons in terms of a weakly interacting
Bose field.}

Above threshold the condensate population becomes macroscopic.
Its growth for increasing $f$ is however suppressed by the
corresponding increase of the population of low energy
excitations. Consequently, the population distribution cannot be
fitted by a Bose-Einstein function. The discrepancy is partly due
to the Bogolubov quasiparticle spectrum -- characterizing an
interacting Bose gas at thermal equilibrium -- and partly to
quantum fluctuations. In order to distinguish the two
contributions, we compare the kinetic result to a distribution
computed for an equilibrium interacting Bose gas in the HFB Popov
limit, \cite{griffin96} accounting for spatial confinement. For
this equilibrium distribution, we assume the same density as
obtained from the kinetic model for a given pump $f$, while the
temperature is extrapolated from the slope of the high-energy
tail of the same kinetic model distribution. In Fig.~
\ref{fig:3}~(a), the result for
$f=50~\mbox{ps}^{-1}~\mu\mbox{m}^{-2 }$, is compared to the
equilibrium HFB Popov distribution with $n=100~\mu\mbox{m}^{-2}$
and $T=20 \mbox{K}$. As expected, the equilibrium result already
deviates from the ideal distribution, due to the modified
spectrum of the interacting system. However, equilibrium and
kinetic results differ significantly in the low-energy region. In
particular, the kinetic model predicts a larger condensate
depletion. The difference is due to the dominant role played by
quantum fluctuations (see also Fig.(\ref{fig:4}) below), whose
amplitude deviates from the equilibrium prediction and has to be
evaluated by means of a proper kinetic treatment like the present
one. Also in the kinetic model, the energy dispersion is modified
by the presence of the condensate because of the two-body
interaction, displaying the linear Bogolubov spectrum of
collective excitations at low momenta,~\cite{steinhauer02} as
shown in Fig.~\ref{fig:3}~(b). Here, the dashed line highlights
the linear part of the spectrum. The plot shows that, even at the highest pump 
intensity, this feature extends over an energy interval smaller than 0.5 meV, well 
within the measured spectral linewidth.\cite{kasprzak06} Samples with a 
significantly smaller polariton linewidth (longer radiative lifetime) would therefore be 
needed, in order to measure this distinctive feature of BEC.
In Fig.~\ref{fig:4} we show the values of the coherent scattering
rates $v^{(k)}_{k,-k}{\mathop{\mathrm Im}\nolimits}
\{\tilde{m}_k\}$ as a function of the energy of the corresponding
states. As expected, they decrease dramatically for increasing
energy and their contribution to the dynamics vanishes for states
outside the coherent region, thus confirming our initial
assumption of separation into two momentum regions. We see also
that, for increasing excitation intensity the values of the
coherent scattering terms increase, resulting in an increased
amplitude of quantum fluctuations.
\begin{figure}[ht]
\includegraphics[width=.43 \textwidth]{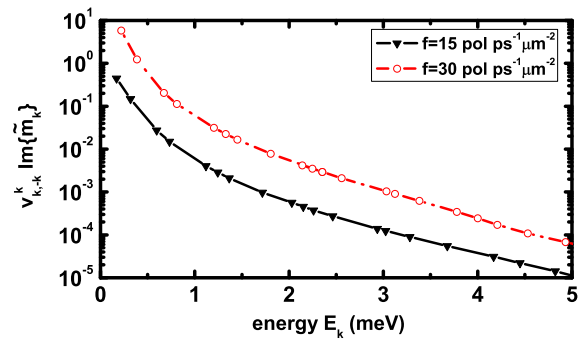}
\caption{Coherent scattering contributions
$v^{(k)}_{k,-k}\mbox{Im} \{\tilde{m}_k\}$,
for two values of the excitation pump $f$. The amplitude of such
processes decreases dramatically for increasing energy and so
their contribution to the dynamics vanishes outside the coherent
region.} \label{fig:4}
\end{figure}

\begin{figure}[ht]
\includegraphics[width=.43 \textwidth]{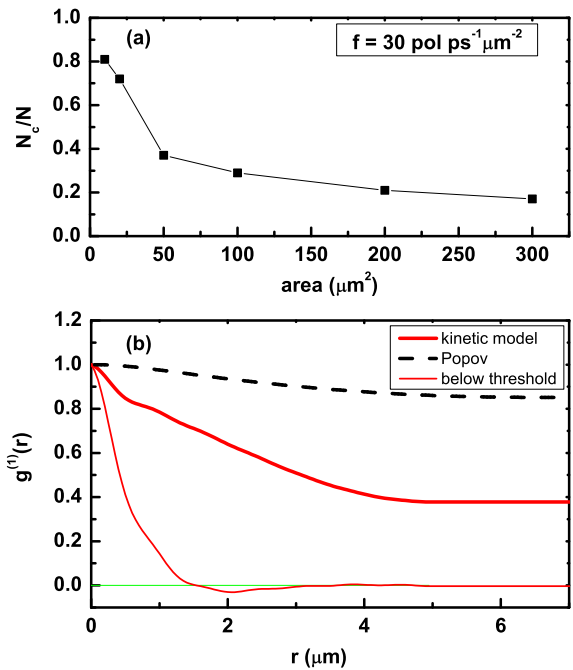}
\caption{(a) Condensate fraction as a function of the system area
for $f=30~\mbox{ps}^{-1}~\mu\mbox{m}^{-2}$. (b) First order
spatial correlation function below and above threshold, compared
to the same quantity resulting from the equilibrium HFB Popov
model.} \label{fig:5}
\end{figure}

For increasing system area $A$, the condensate fraction in the
steady-state regime decreases, as shown in Fig.~\ref{fig:5}~(a),
because coherent scattering is favored by a smaller energy gap
$\Delta$. Thermal and quantum fluctuations will eventually
dominate in the thermodynamic limit of infinite size, resulting
in a full condensate depletion, as required by the
Hohenberg-Mermin-Wagner theorem. Polariton condensation occurs
thanks to the locally discrete nature of the energy spectrum,
induced either by artificial confinement or by disorder. In a
realistic system \cite{richard05,LangbeinPRL02,EldaifAPL2006},
localization could therefore affect the polariton BEC,
independently of other parameters like Rabi splitting and exciton
saturation density.

We finally study the influence of quantum fluctuations on the
one-body spatial correlation $g^{(1)}({\bf r},{\bf
r}^\prime)=n({\bf r},{\bf r}^\prime)/\sqrt{n({\bf r})n({\bf
r}^\prime)}$. The one-body density matrix $n({\bf r},{\bf
r}^\prime)$ is the direct expression of ODLRO that characterizes
BEC.~\cite{bloch00,penrose56} It depends on $|{\bf r}-{\bf
r}^\prime|$ for a uniform system and can be computed in terms of
the Fourier transform of the population $N_k$. The density in the
denominator renormalize the shape of the condensate wave
function, hence we expect the averaged quantity $g^{(1)}({\bf
r})\equiv 1/A\int d{\bf R} g^{(1)}({\bf R},{\bf R}+{\bf r})$ to
be scarcely affected by the assumption of a uniform condensate.
In Fig.~\ref{fig:5}~(b), we plot $g^{(1)}({\bf r})$ below and
above the condensation threshold. Below threshold, correlations
vanish for distances larger than $1-2~\mu\mbox{m}$, as predicted
by both the kinetic model and the equilibrium HFB Popov approach.
Above threshold, the correlation length increases and
$g^{(1)}(r)$ remains finite over the whole system size. However,
for all values of the pump, $g^{(1)}(r)$ remains smaller than
$0.5$ at distances larger than $5~\mu\mbox{m}$, whereas the
equilibrium HFB Popov result is significantly larger, due to a
larger condensate fraction. Therefore the effect of quantum
fluctuations turns out to strongly affect the formation of ODLRO.
The quantity $g^{(1)}(r)$ could be easily accessed in an
experiment in which the light emitted by different positions on
the sample is made interfere. For such an experiment, we
therefore predict the increase of the spatial correlation length
as a signature of condensation, with a correlation staying below
0.5 because of quantum fluctuations, even far above the threshold.

{Finally, we mention that we have also performed simulations for material parameters modeling a GaAs microcavity
with Rabi splitting $2\hbar\Omega_{R}=7~\mbox{meV}$, detuning $\delta=0~\mbox{meV}$, lattice
temperature $T=10~\mbox{K}$, and system area $A=100~\mu\mbox{m}^2$. In this
case $v_{XX}=6\times 10^{-5}~\mbox{meV}$, $v_{sat}=-0.15\times
10^{-5}~\mbox{meV}$ and the resulting polariton-polariton
interaction matrix element at zero momentum is $v^{(0)}_{0,0}=1.5\times
10^{-5}~\mbox{meV}$. For this case as well, we observe the
occurrence of polariton condensation and the partial suppression
of the ODLRO (not shown).
Quantitatively, we notice that in this case the total exciton
density at threshold is $n_x\simeq 500~\mu\mbox{m}^{-2}$, significantly higher than in the previous case. Nevertheless, as the number of quantum wells needed to achieve
$7~\mbox{meV}$ Rabi splitting is at least $4$, the estimated exciton density
per quantum well is about $100~\mu\mbox{m}^{-2}$, namely still one order of magnitude lower than the saturation
density for GaAs, $n_{sat}=2\times10^{3}~\mu\mbox{m}^{-2}$.\cite{Schmitt-Rink1985}
Our approach is thus justified also in this case.}

\section{Conclusions}
In conclusion, the present model shows that the dynamics of
quantum fluctuations significantly affects polariton BEC and the
formation of ODLRO in a polariton condensate. In a typical case,
quantum fluctuations partially deplete the condensate, already
slightly above threshold. Quantitatively, the effect depends on
the locally discrete energy spectrum, due to trapping or to
disorder. We predict that the observation of BEC
and ODLRO should be favored by smaller polariton size, as in the
recently studied polariton ``quantum boxes'',\cite{EldaifAPL2006}
or in local minima of the disorder potential. This suggests that,
for a given sample, a study of the polariton localization length
in the lowest energy states~\cite{LangbeinPRL02} could give
deeper insight into the BEC mechanism.

\begin{acknowledgments}
We are grateful to I. Carusotto and R. Zimmermann for fruitful
discussions. We acknowledge financial support from the Swiss
National Foundation through project N. PP002-110640.
\end{acknowledgments}

\appendix*

\section{}

Equations (\ref{eq:tot}) are derived under the assumption that 3-body and higher order
correlation terms can be factored. In this Appendix we provide some detail on their derivation. Equations for the populations $N_c$ and $\tilde{N}_k$, with respect to the 
polariton-polariton interaction part, result directly from Heisenberg time-evolution of operators. The corresponding exciton-phonon term $\dot{N}|_{ph}$ is obtained within 
a standard Boltzmann kinetics in the Markov limit,\cite{doan05}, while for the exciton-exciton term $\dot{N}|_{XX}$ we have used the equations introduced by Porras 
{\em et al.} in Ref. \onlinecite{porras02}. Some less straightforward steps were taken to obtain the equations describing 
the scattering
amplitudes $\tilde{m}_k$. First, as pointed out in the text, we included only the dynamics induced by polariton-polariton scattering Hamiltonian within the coherent region. 
Then, Heisenberg equations of motion result in the following equations
\begin{eqnarray}
i\hbar \dot{\tilde{m}}_k &=& 2(\hbar \omega_k - v^{(0)}_{k,0})
\tilde{m}_k +
v^{(0)}_{k,0}\langle \hat{a}^{\dagger} \hat{a}^{\dagger} (\hat{a}^{\dagger} \hat{a} + \hat{a} \hat{a}^{\dagger}) \tilde{p}_k \tilde{p}_{-k} \rangle \nonumber \\
&-& 4 \sum_q v^{(0)}_{k,0} \langle \hat{a}^{\dagger} \hat{a}^{\dagger} \tilde{p}_q^{\dagger} \tilde{p}_{q} \tilde{p}_k \tilde{p}_{-k} \rangle \nonumber \\
&+& v^{(k)}_{k,-k} \langle \hat{a}^{\dagger} \hat{a}^{\dagger} \hat{a} \hat{a} (\tilde{p}_{-k}^{\dagger} \tilde{p}_{-k} + \tilde{p}_k \tilde{p}_{k}^{\dagger}) \rangle \nonumber \\
&+& \sum_{qq'} v^{(k-q)}_{q,q'}\langle \hat{a}^{\dagger}
\hat{a}^{\dagger} (\tilde{p}_{q+q'-k}^{\dagger} \tilde{p}_{-k} +
\tilde{p}_k \tilde{p}_{q+q'+k}^{\dagger}) \tilde{p}_q
\tilde{p}_{q'}\rangle \nonumber \\
&-& \sum_q v^{(q)}_{q,-q}
\langle (\hat{a} \hat{a}^{\dagger} +
\hat{a}^{\dagger}\hat{a})\tilde{p}_q^{\dagger}
\tilde{p}_{-q}^{\dagger} \tilde{p}_k \tilde{p}_{-k} \rangle.
\end{eqnarray}
Then, by neglecting 3-body correlations, we can introduce the following
factorizations:
\begin{eqnarray}
&&\langle\hat{a}^{\dagger} \hat{a}^{\dagger} \hat{a}^{\dagger}
\hat{a} \tilde{p}_k \tilde{p}_{-k}\rangle \simeq (N_{c} - 2)
\tilde{m}_k; \\
&&\langle\hat{a}^{\dagger}\hat{a}^{\dagger}\tilde{p}_q^{\dagger}\tilde{p}_{q}\tilde{p}_k\tilde{p}_{-k}
\rangle  \simeq (\tilde{N}_q-\delta_{q,k}
\tilde{N}_k-\delta_{q,-k}\tilde{N}_{-k})\tilde{m}_k \nonumber \\
&&\hspace{2.6cm}+\tilde{N}_{q,k}\tilde{m}_{q,-k}+\tilde{N}_{q,-k}\tilde{m}_{q,k};\\
&&\langle \hat{a}^{\dagger}\hat{a} \tilde{p}_q^{\dagger}
\tilde{p}_{-q}^{\dagger} \tilde{p}_k \tilde{p}_{-k} \rangle
\simeq N_{c} \langle \tilde{p}_q^{\dagger}
\tilde{p}_{-q}^{\dagger} \tilde{p}_k \tilde{p}_{-k} \rangle;\\
&&\langle \hat{a}^{\dagger} \hat{a}^{\dagger} \hat{a} \hat{a}
\tilde{p}_{k}^{\dagger} \tilde{p}_{k}  \rangle \simeq
N_{c}(N_{c}-1)\tilde{N}_k;\\
&&\langle \hat{a}^{\dagger} \hat{a}^{\dagger}
\tilde{p}_{q+q'-k}^{\dagger} \tilde{p}_{-k} \tilde{p}_q
\tilde{p}_{q'}\rangle \simeq 2\tilde{N}_{q+q'-k,q'}
\tilde{m}_{q,-k} \nonumber \\
&& \hspace{3.5cm}+\tilde{N}_{q+q'-k,-k} \tilde{m}_{q,q'} \nonumber \\
&&\hspace{3.5cm}- \delta_{q,k}\delta_{q',k} \tilde{N}_k
\tilde{m}_k).
\end{eqnarray}
From here, within the assumption of a spatially homogeneous
system, som of the momentum sums can be carried out explicitely and Eq.(\ref{eq:tot}) is finally obtained.

We now turn to the evaluation of the two-body correlations $\langle
\tilde{p}_q^{\dagger}\tilde{p}_{-q}^{\dagger}\tilde{p}_k\tilde{p}_{-k}
\rangle$.
As explained in the text, they are evaluated in a quasi-stationary limit, via the stationary solution of the Bogolubov problem. In terms of collective excitation operators
$\hat{\Lambda}_k$, defined in Eq. (\ref{eq:bogo}), this quantity is rewritten as:
\begin{eqnarray}
\langle
\tilde{p}_q^{\dagger}\tilde{p}_{-q}^{\dagger}\tilde{p}_k\tilde{p}_{-k}
\rangle &\simeq& \frac{\langle
\hat{a}\hat{a}\hat{a}^{\dagger}\hat{a}^{\dagger}\tilde{p}_q^{\dagger}\tilde{p}_{-q}^{\dagger}\tilde{p}_k\tilde{p}_{-k}
\rangle}{(N_c+1)(N_c+2)} \\
&=& \frac{N^2}{(N_c+1)(N_c+2)}\langle
\hat{\Lambda}_q^{\dagger}\hat{\Lambda}_{-q}^{\dagger}\hat{\Lambda}_k\hat{\Lambda}_{-k}
\rangle. \nonumber
\end{eqnarray}
The correlation amplitude for the field $\hat{\Lambda}_k$ is then computed by
using the Bogolubov transformation
$\hat{\Lambda}_k=U_k\hat{\alpha}_k+V^*_{-k}\hat{\alpha}^{\dagger}_{-k}$.
All the resulting terms are factored as products of the Bogolubov
quasi-particle populations $\bar{N}_k=\langle
\hat{\alpha}^{\dagger}_k \hat{\alpha}_k \rangle$, as:
\begin{equation}
\langle \hat{\alpha}^{\dagger}_k \hat{\alpha}^{\dagger}_{q}
\hat{\alpha}_{k}\hat{\alpha}_{q}\rangle \simeq
\bar{N}_k\left(\bar{N}_q-\delta_{kq}\right).
\end{equation}
Collecting all the terms, we finally recover
Eq.(\ref{eq:2corr_bog}). In this equation, single-particle and quasi-particle populations are related with each other through the expression
\begin{equation}
\langle \tilde{p}^{\dagger}_k \tilde{p}_{k} \rangle \simeq
\frac{N}{N_c+1} [(|U_k|^2 + |V_k|^2)\bar{N}_k + |V_k|^2].
\end{equation}


\begin{thebibliography}{10}
\bibitem{note} The restriction to the lower polariton branch implies the
approximation of structureless polaritons. We
assume that the Rabi splitting and the Hopfield factors do not
change within the range of densities considered. This assumption
was checked evaluating the density-renormalized Rabi splitting
$\tilde{\Omega}_R=\Omega_R(1-N_{exc}/N_{sat})$.\cite{ciuti03}
\end{thebibliography}
\end{document}